\documentclass[epj]{svjour}
\usepackage{epsfig}
\begin{document}
\title{HERA results on the structure of the proton\\{\small\rm (Invited talk given at Quark and Nuclear Physics 2002, Juelich, June, 2002)} }
\author{B. Foster\inst{1,2}%
}                     
\institute{H.H. Wills Physics Laboratory, University of Bristol, Bristol, U.K. \and DESY, Notkestrasse 85, Hamburg, Germany}
\date{4th November, 2002}
\abstract{A selection of results from the H1 and ZEUS experiments at HERA are reviewed, particularly in the area of deep inelastic scattering and diffraction. Quantum chromodynamics gives a good explanation of these data down to surprisingly low values of the four-momentum transfer, $Q^2$. Data at smaller
$Q^2$ can be described by Regge models as well as by dipole models including
parton-saturation effects. The latter can also give a unified description
of many features of diffractive data.
\PACS{13.10.+q
         \and
      13.60.-r 
     } 
} 
\maketitle
\section{Introduction}
\label{sec:intro}
Results from HERA have revolutionised our view of the structure of the proton.
The increased kinematic range has not only extended the range upwards in
$Q^2$, the negative of the squared four-momentum transfer to the proton, but also downwards in $x$, the fractional momentum carried by the struck parton. ZEUS
data reaches down to $x \sim 10^{-6}$. The rapid rise of the gluon density
as $x$ decreases, discovered at HERA, implies that the density is
very high at these values of $x$, so that it is interesting to see if
non-linear effects in gluon
evolution, often referred to as saturation, become visible. 
Other deviations from the standard DGLAP evolution in quantum chromodynamics 
(QCD) may also
become visible in specific kinematic regions, particularly at low
$x$. Here, there is a transition from the QCD description
that works exceedingly well at high $Q^2$ to a Regge description that
is applicable at low $Q^2$. These features are illustrated with reference
to next-to-leading-order QCD fits carried out both by H1 and ZEUS. The
description of these phenomena by dipole models is discussed. These form a natural way to explore the relationship between inclusive deep inelastic scattering and diffractive interactions. 
\section{The $F_2$ structure function}
\label{sec:F2}
The double-differential cross section, $d^2\sigma/dQ^2dx$, for the deep inelastic scattering (DIS) of an unpolarised positron from a proton can be described by
\begin{eqnarray}  
\frac{d^2 \sigma} 
{dx dQ^2} & = & \frac{2\pi \alpha^2}{x Q^4}   
 \left[   
Y_+ \cdot
 F_2(x, Q^2) \right. \nonumber \\
 & \; - & \left.  {y^2} F_L(x, Q^2) - Y_- \cdot x F_3(x, Q^2) 
\right] 
\label{eq:Fl:sigma} 
\end{eqnarray}
where $Y_{\pm}$ are kinematic factors given by 
\begin{equation}
Y_{\pm} = 1 \pm (1-y)^2
\label{eq:Y}
\end{equation}
$y = Q^2/sx$, $s$ is the total squared centre-of-mass energy
and $F_2, F_L$ and $F_3$ are structure functions. At low $Q^2$,
the effects of $xF_3$ can be ignored and, since the effects of $F_L$
are small except at high $y$, the double-differential cross section
is approximately proportional to $F_2$. The measurement of the differential
cross section therefore gives information on $F_2$, which in its turn can
be decomposed (in the DIS normalisation scheme) into    
\begin{equation} 
 F_2(x, Q^2)  =  \sum_{i=u,d,s,c,b} A_i(Q^2) \left[ xq_i(x,Q^2) + 
   x\overline{q}_i(x,Q^2) \right].  
\label{eq:F2:qpm}  
\end{equation} 
The parton densities $q_i(x,Q^2)$ and $\overline{q}_i(x,Q^2)$ refer  
to quarks and antiquarks of type 
$i$. At low $Q^2$, the quantities
$A_i(Q^2)$ are given by the square of the electric charge of quark or 
antiquark $i$. Figure~\ref{fig:1} shows the $F_2$ structure function 
measured by the ZEUS and H1 
collaborations\cite{H1F2-9697a,H1F2-9697b,ZEUSF2-9697} at high $x$,
together with data from the fixed-target experiments NMC\cite{np:b483:3},
BCDMS\cite{pl:b223:485,pl:b237:592} and E665\cite{pr:d54:3006} as a function
of $Q^2$ in bins of $x$. The region of scaling around $x=0.2$ can be
seen clearly, together with the strong scale breaking as $x$ falls, caused by
gluon radiation. The fit to the predictions of next-to-leading-order (NLO) 
QCD{\cite{H1F2-9697b} can be seen to give an excellent fit to the data over the whole kinematic range. 
\begin{figure}[h]
\begin{center}
\epsfig{file=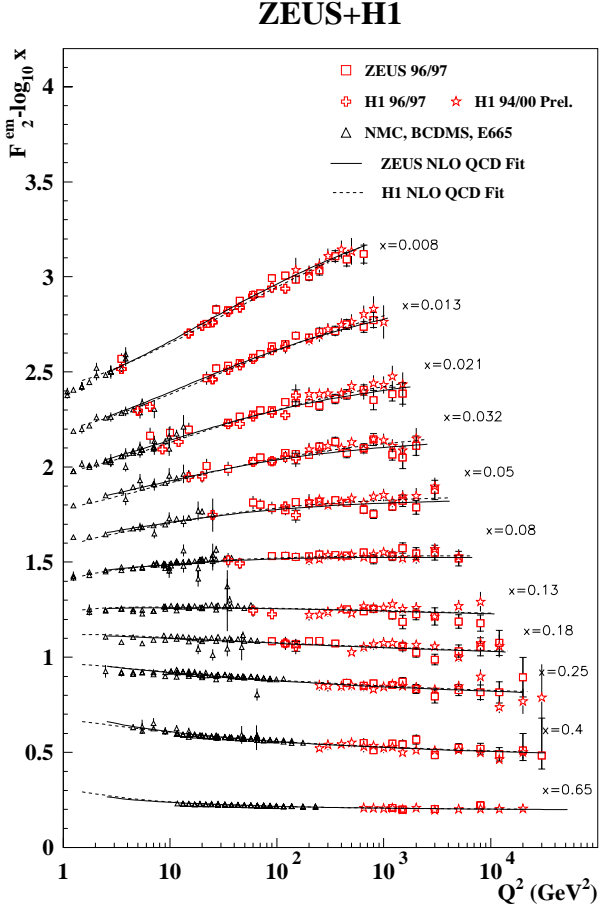,%
      width=10cm,%
      height=14cm,%
      clip,%
       }
\end{center}
\caption{The H1 and ZEUS data on $F_2$ at high $x$ as a function of $Q^2$. Also shown are points from the fixed-target experiments NMC, BDCMS and E665. The solid curve shows the NLO QCD fit. Each bin in $x$ is displaced by the factor 
$\log_{10} x$ as indicated for ease of visibility.}
\label{fig:1}
\end{figure}
ZEUS has performed an NLO QCD fit which
is also of excellent quality\cite{ZEUS-NLOpaper}. Both experiments have used the results of the fit to determine the parton density distributions through relations analogous to  
that given by Eq.~\ref{eq:F2:qpm}. Some of the results from ZEUS are shown in 
\begin{figure}[h]
\begin{center}
\epsfig{file=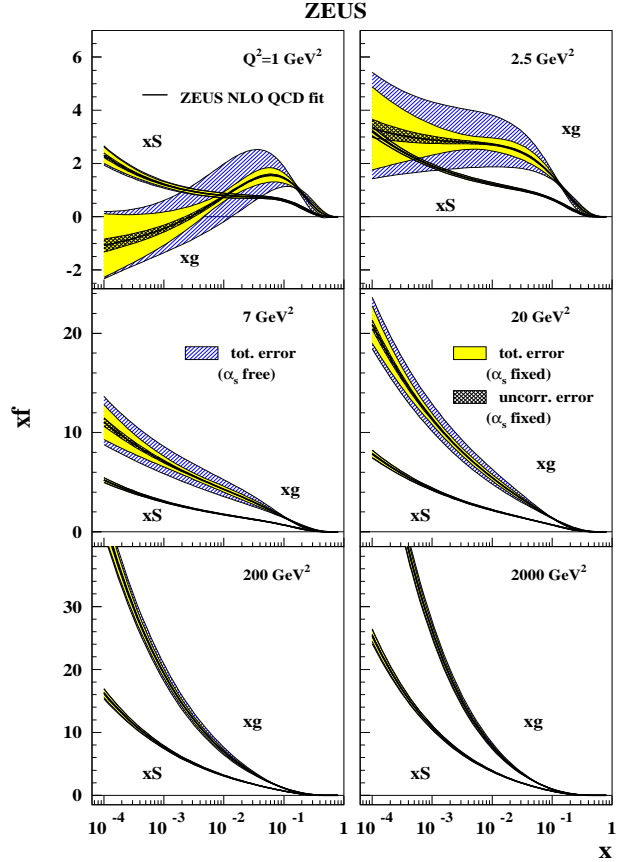,%
      width=10cm,%
      height=12cm,%
      clip%
        }
\end{center}
\caption{The ZEUS determination of the gluon and singlet-quark densities
in six bins of $Q^2$ as a function of $\ln x$. The shaded bands show the
uncorrelated error on the parton density, the total error if the strong coupling is fixed, and the total error if $\alpha_s$ is allowed to vary.}
\label{fig:2}
\end{figure}
Fig.~\ref{fig:2}.

Several conclusions can be drawn from these sorts of determinations. Firstly, the strong rise of the gluon density as $x$ falls at medium and high $Q^2$ can
clearly be seen. The singlet quark densities have a lower density than that of the gluon, as would be expected since the sea is considered to be generated by the emission of gluons from the valance quarks. However, at $Q^2 \sim 1$ GeV$^2$ and at low $x$, the gluon density drops considerably below that of the sea
and, even more strikingly, seems to become negative. Although worrying and difficult to interpret, this behaviour is not necessarily pathological since parton densities are not observables; however, at around the same $Q^2$, $F_L$ also seems to become negative. The conclusion is, therefore, that the NLO QCD 
formalism begins to break down at $Q^2 \sim 1$ GeV$^2$. In fact, the biggest
surprise is probably that QCD and the concept of partons 
does indeed seem to work down to such low values of $Q^2$.
\section{$F_2$ at low $Q^2$}
\label{sec:F2lowQ2}
For values of $Q^2$ less than 1 GeV$^2$, a good representation of the
data can be achieved using Regge theory, particularly at low $x$,
since this corresponds to high $W$, where $W$ is the centre-of-mass energy
of the photon-proton system, at which Regge theory is most
applicable. This can be seen in Fig.~\ref{fig:3}, which shows the
high-precision ZEUS BPT $F_2$ data\cite{ZEUS-BPT} 
at low $Q^2$, together with other H1 and ZEUS data,
as a function of $y$ versus $\log Q^2$.
For $Q^2 > 1$ GeV$^2$, the data
are approximately flat, consistent with the expectations from QCD. Below
1 GeV$^2$, the data fall rapidly, consistent with the $Q^{-2}$ dependence expected from conservation of the electromagnetic current. The form is well
fitted by a parameterisation\cite{ZEUS-BPT}
based on Regge Theory.
\begin{figure}[h]
\begin{center}
\epsfig{file=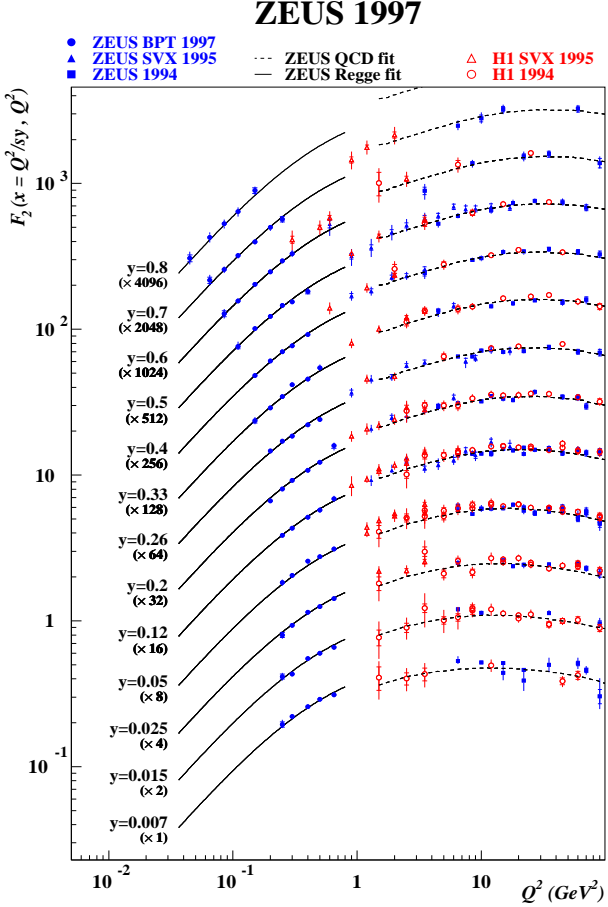,%
      width=8cm,%
      height=12cm,%
      clip%
        }
\end{center}
\caption{H1 and ZEUS $F_2$ data as a function of $\log Q^2$. Each $y$ bin is multiplied by the number in parentheses for clarity of presentation. For $Q^2 > 1$ GeV$^2$, the curves show an NLO QCD fit to the data. For the ZEUS data
below 1 GeV$^2$, a fit to a Regge-based parameterisation is shown.}
\label{fig:3}
\end{figure}
These features are confirmed by recent H1 data taken with the interaction vertex shifted to give access to smaller values of $Q^2$. These data\cite{H1-SVX2000} cover the
$Q^2$ region between the ZEUS BPT data and that taken in the central detectors with $Q^2 > 2 - 3$ GeV$^2$.
\subsection{Low $x$ phenomenology}
\label{sec:lowx}
We have seen that the QCD description of the DIS data begins to break down
around $Q^2 \sim 1$ GeV$^2$, and that for values below this, a good description can  be given by Regge theory. Neither theory can describe the entire HERA kinematic range. One possibility to achieve this is via dipole models, which consider the DIS process in the rest frame, rather than the more usual infinite-momentum frame, of the proton. In such a frame, the virtual photon can be considered to fluctuate into partonic subsyetems, typically a quark-antiquark pair, or dipole, or a quark-antiquark-gluon state. By combining such a treatment with the concept of parton saturation, models such as that of 
Golec-Biernat and W\"{u}sthoff\cite{pr:d59:014017,pr:d60:114023} 
can give a good description of the data over all ranges of $Q^2$. 
Parton-saturation effects would be expected to set in at low $x$ when the parton density becomes so large that their probability of interacting with each other becomes non-negligible. Fits of the Golec-Biernat and W\"{u}sthoff model to the HERA data, including one modified to take account of QCD-evolution
effects\cite{GBW-evolution}, 
is shown in Fig.~\ref{fig:4} for the higher-$Q^2$ data. 
The fit is significantly improved by including QCD evolution. 
The model also gives a good fit to the ZEUS BPT data (not shown).
\begin{figure}[h]
\begin{center}
\epsfig{file=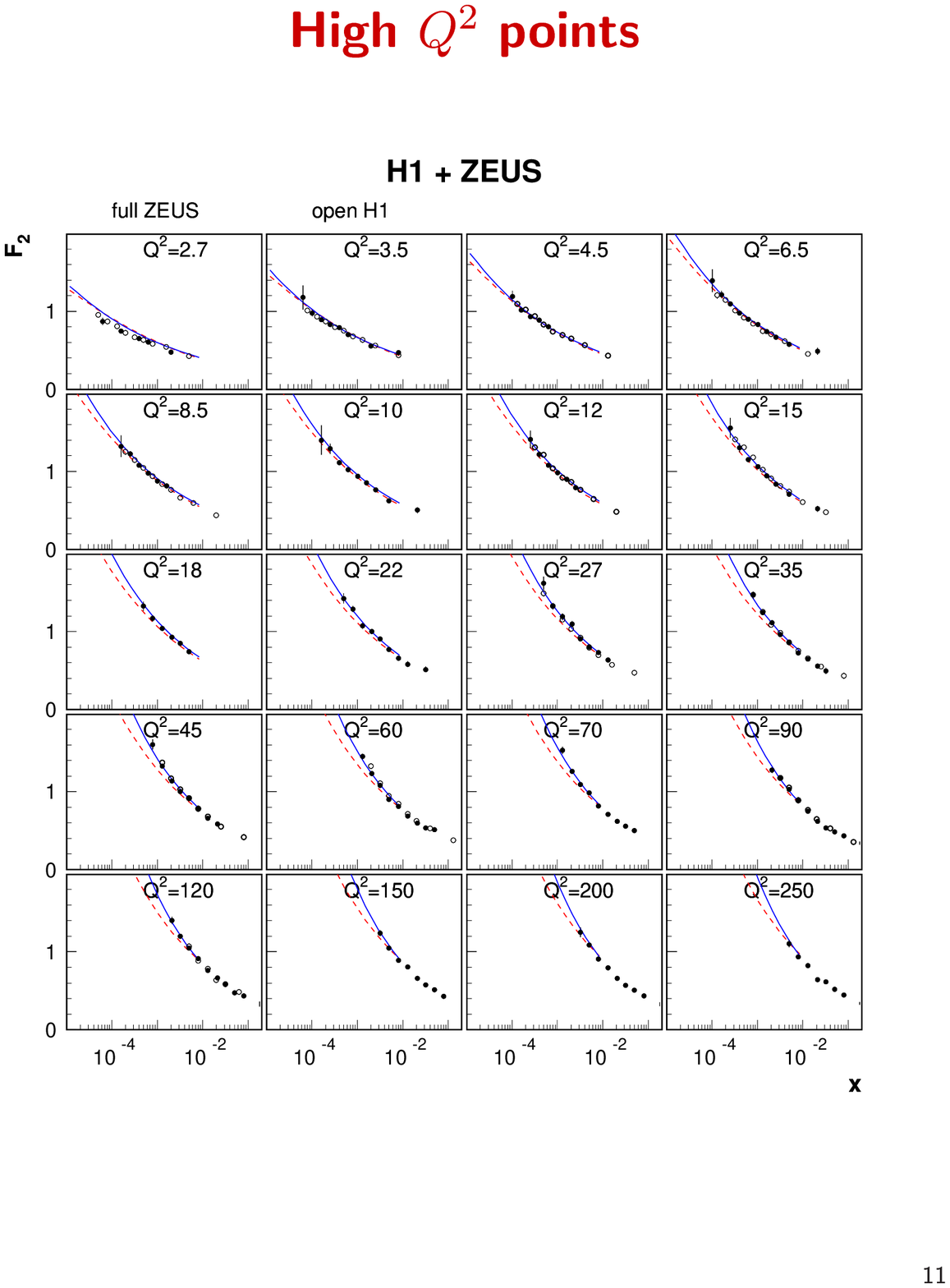,%
      width=8cm,%
      height=12cm,%
      clip%
        }
\end{center}
\caption{The ZEUS (full points) and H1 (open circles) $F_2$ data as a function of $x$ in bins of $Q^2$. Also shown are two fits to the Golec-Biernat and W\"{u}sthoff model. The dotted line shows the fit to the original model without QCD evolution; the full curve shows the prediction taking such evolution into account.} 
\label{fig:4}
\end{figure}
\section{Diffraction}
\label{sec-diffraction}
Dipole models not only successfully fit the DIS data over the full HERA kinematic range, but also give a natural connection to another important process, diffraction. In diffractive processes at HERA, a hard interaction occurs but nevertheless the proton leaves the interaction intact. This can be understood if the proton exchanges a colour-singlet object which transfers energy and momentum to the interaction with the virtual photon. This object is normally called the Pomeron; it can be considered to have a partonic structure analogous to that of the proton, which can be investigated by examining the debris of the hard collision, as is done in the measurement of the proton structure functions.    
The connection between this process and standard inclusive DIS in dipole models is that the inclusive DIS processes can be considered to be the sum of 
single-gluon and two-gluon exchange between the proton and the dipole fluctuation of the virtual photon, whereas diffraction can be considered to originate from colour-singlet two-gluon exchange only. Such a picture implies a simple connection between two apparently rather different processes. 

The success of the dipole picture can be seen in Fig.~\ref{fig:5}, which shows the ratio of the diffractive cross section to the total cross section as a function of $W$ in various $Q^2$ and $M_X$ bins, where $M_X$ is the mass of the diffractive system. The very flat behaviour of this ratio as a function
of $W$ is in contradiction to simple ideas from Regge theory and the optical theorem; the former would predict a $W$ dependence of diffraction like $W^{0.16}$ compared to that of the total cross section of $W^{0.4}$. Similarly, if diffraction is related to the exchange of a colour-singlet two-gluon system, then it should have a $W \sim 1/x$ dependence like the gluon-density squared, compared to the single-gluon dependence of the inclusive cross section, which
ought to result in the ratio strongly rising as $W$ increases. The flatness of this ratio is therefore a surprise, but, at least for the smallest values of $M_X$, it is rather well fitted by the Golec-Biernat and W\"{u}sthoff model.           
\begin{figure}[h]
\begin{center}
\epsfig{file=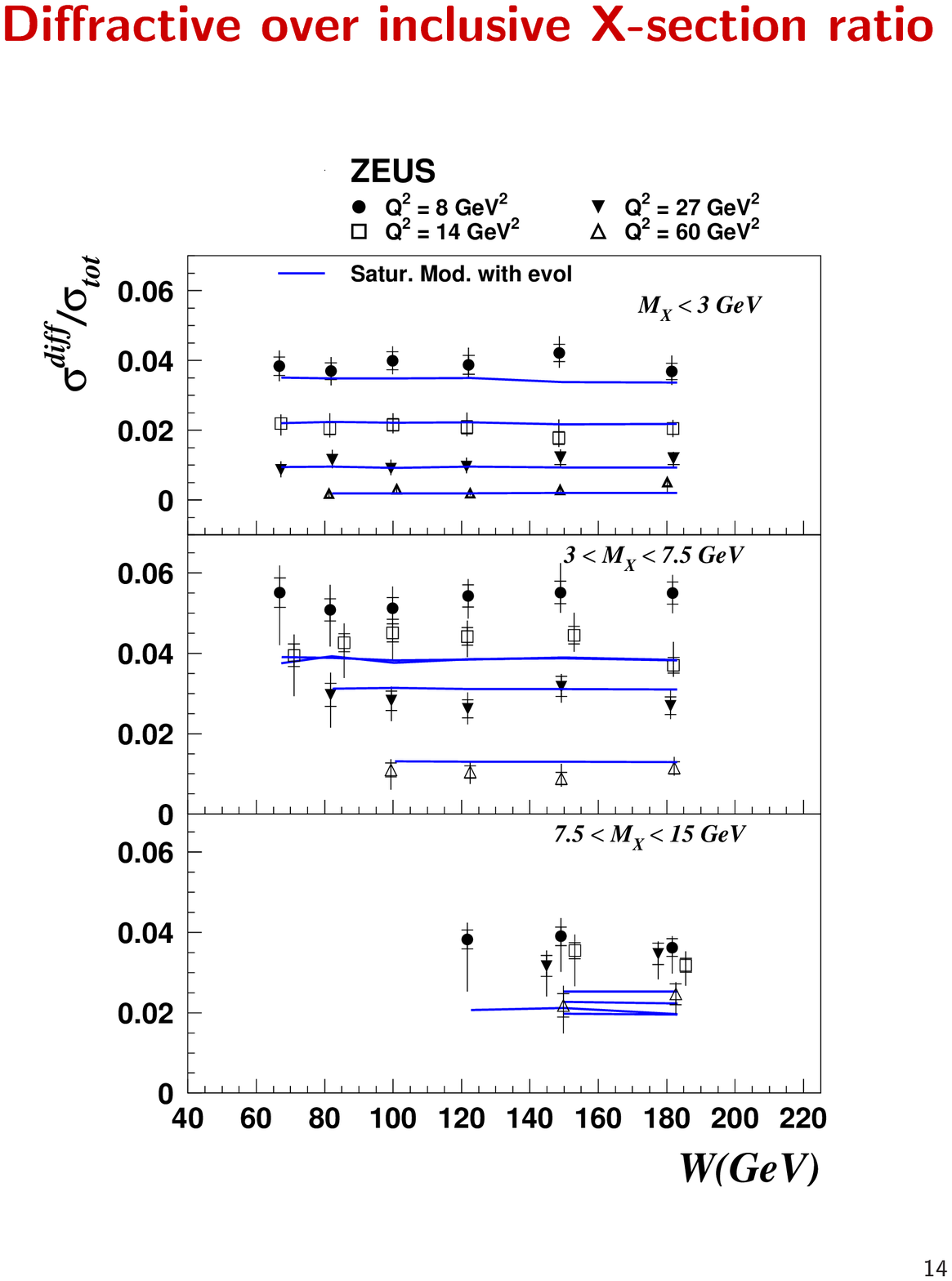,%
      width=8cm,%
      height=11cm,%
      clip%
        }
\end{center}
\caption{The ratio of the diffractive cross section to the total cross section as a function of $W$ in four bins of $Q^2$ and in three regions on $M_X$. The curve shows the predictions of the Golec-Biernat and W\"{u}sthoff model.}  
\label{fig:5}
\end{figure}

A similar picture emerges if one examines more exclusive processes, such as elastic vector-meson production\cite{Amsterdam820}. Here the light mesons, $\rho$ and $\phi$, show a similarly flat cross-section ratio. In the case of the $J/\psi$, however, where the mass of the charmed quark is sufficiently large to form a hard scale, the behaviour of the ratio in photoproduction is very different, with a steep dependence on W which flattens at higher $Q^2$. This is a clear indication of the interplay of hard scales in QCD and the study of such processes is likely to be very interesting. 

The study of vector-meson production as a function of $t$, the momentum transfer at the proton vertex, is also interesting\cite{ZEUS-hight}. There is considerable evidence that $t$ can also form a hard scale in QCD. Forshaw and Poludniowski\cite{Jeff} have looked at the production rate of $\rho, \phi$ and $J/\psi$ mesons as a function
of $t$ and find that models based on conventional two-gluon exchange cannot fit the data. Their model, based on a BKFL-like\cite{BFKL} Pomeron, can fit the data over the full $t$ range for all mesons. This is one of the few concrete indications of the importance of BFKL evolution at HERA.

The H1 collaboration has recently produced\cite{H1-Pomeronpdf} a determination of the parton content of the Pomeron from a NLOQCD fit to the reduced cross sections obtained from the inclusive diffractive data, 
analogous to that discussed above for the $F_2$ data. 
The results of this fit
for the singlet and gluon distributions in the Pomeron are shown in 
Fig.~\ref{fig:6} as a function of $z$, the fractional momentum of the Pomeron
carried by the parton.
\begin{figure}[h]
\begin{center}
\epsfig{file=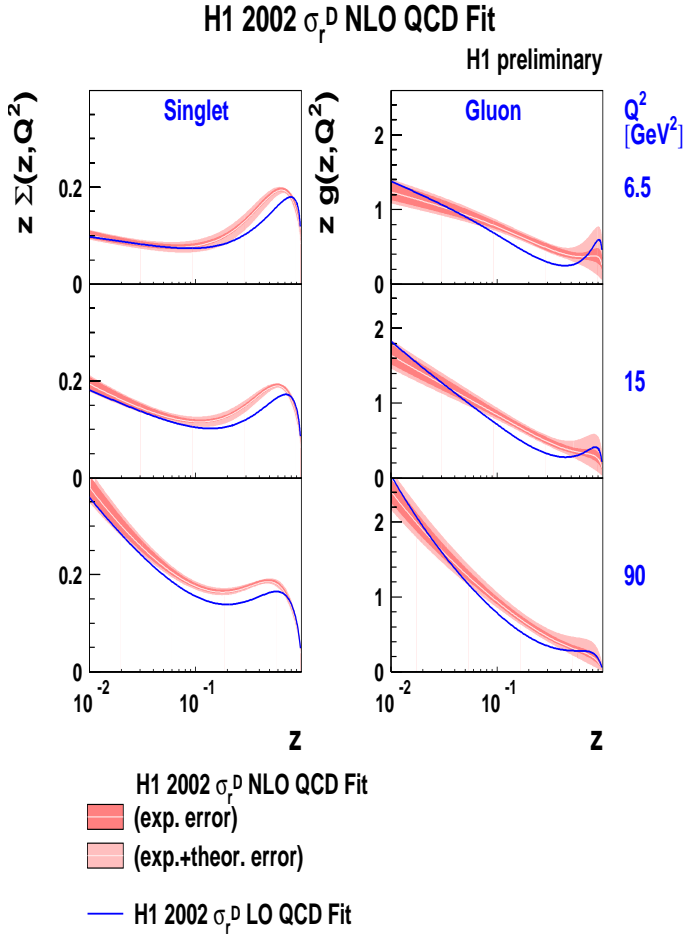,%
      width=9cm,%
      height=13cm%
        }
\end{center}
\caption{The singlet and gluon parton distribution functions of the Pomeron as obtained from the H1 reduced diffractive cross sections as functions of $z$, the
fractional momentum of the Pomeron carried by the parton. The distributions
are shown in three bins of $Q^2$. The line shows the LO QCD fit, while the
shaded bands indicate the results of the NLOQCD fit with uncertainties as indicated in the legend.}  
\label{fig:6}
\end{figure}
The fits clearly show that the Pomeron is dominated by gluons and that, similarly to the proton, the gluon distribution rises strongly as $z$ falls, driving in turn a rise in the singlet distribution at higher $Q^2$.  
These data are in agreement with the idea that diffraction is mediated by
an exchange of some form of two-gluon ladder.

Finally in this section, although not strictly diffractive, it is instructive
to consider the data on the pion structure function. Virtual pions can be
probed in $ep$ DIS by tagging leading neutrons in a dedicated calorimeter at
$0^{\rm o}$ far downstream of the interaction point in the proton direction. Such
a process can be shown in particular kinematic regions to be dominated by exchange of another colourless object, the virtual pion. Thus, similarly to
the H1 determination of the Pomeron pdfs discussed above, ZEUS has used the
cross sections for inclusive leading neutron production to determine the
pion structure function\cite{ZEUS-pionstructure}. The data are
generally compatible with the factorisation of the lepton and proton vertices,
although a factorisation breaking with $Q^2$ of around 20\% is observed, which
may be caused by absorptive rescattering of the neutron. The overall normalisation of the pion structure function derived from these data 
is uncertain to around a factor of two due to the lack of an unambiguous model for the virtual-pion flux. Nevertheless, the shape of the pion structure in the
accessible kinematic region is remarkably similar to that of the proton. 
         
\section{Summary}
Within the limited space of this write-up, it is not possible to cover a vast field of other HERA physics, in particular that at high $Q^2$ and the search
for physics beyond the Standard Model of particle physics. This area will
be the focus of the HERA II programme, which will not only benefit from
a substantial upgrade in the luminosity, but also from the availability of
polarised lepton beams. Thus, the field of heavy-quark physics, electroweak 
studies and exotic physics will take centre stage at HERA II. The richness
of the HERA I data, particularly in the areas of the low-$Q^2$ interface,
diffraction and jet physics are far from exhausted, however, and the next
few years should bring further illumination in this poorly understood and
theoretically challenging area of physics.  

\section*{Acknowledgements}
I am grateful to many colleagues, particularly on ZEUS and H1, for helpful discussions. I am particularly grateful to M. Kuze for a critical reading of this manuscript.

\end{document}